\documentclass[10pt,twocolumn,letterpaper]{article}

\usepackage[pagenumbers]{cvpr} 

\usepackage[T1]{fontenc}
\usepackage{csquotes}
\usepackage{fancyhdr}
\usepackage{listings}
\usepackage{placeins}

\MakeOuterQuote{"}

\fancypagestyle{firstpage}{%
  \fancyhf{}%
  \fancyhead[L]{\parbox{\textwidth}{\scriptsize%
    \copyright~2026 IEEE. Personal use of this material is permitted.
    Permission from IEEE must be obtained for all other uses, in any
    current or future media, including reprinting/republishing this
    material for advertising or promotional purposes, creating new
    collective works, for resale or redistribution to servers or lists,
    or reuse of any copyrighted component of this work in other works.}}%
}

\definecolor{backcolor}{HTML}{F2F2F2}
\definecolor{codecomment}{HTML}{0393C4}
\definecolor{codekeyword}{HTML}{DC307F}
\definecolor{codestring}{HTML}{30BB30}
\lstdefinestyle{mystyle}{
    backgroundcolor=\color{backcolor},
    commentstyle=\color{codecomment},
    keywordstyle=\color{codekeyword},
    stringstyle=\color{codestring},
    basicstyle=\ttfamily\scriptsize,
    breakatwhitespace=false,
    breaklines=true,
    captionpos=b,
    keepspaces=true,
    showspaces=false,
    showstringspaces=false,
    showtabs=false,
    tabsize=4,
    framexleftmargin=10pt,
    framextopmargin=8pt,
    framexbottommargin=8pt,
    frame=tb,
    framerule=0pt,
}
\definecolor{cvprblue}{rgb}{0.21,0.49,0.74}
\usepackage[pagebackref,breaklinks,colorlinks,allcolors=cvprblue]{hyperref}

\usepackage{tabularx}

\def\paperID{} 
\def\confName{CVPR}
\def\confYear{2026}

\title{DeDelayed: Deleting Remote Inference Delay via On-Device Correction}

\author{
    Dan Jacobellis\textsuperscript{1,2},
    Mateen Ulhaq\textsuperscript{2},
    Fabien Racap\'e\textsuperscript{2},
    Hyomin Choi\textsuperscript{2},
    Neeraja~J.~Yadwadkar\textsuperscript{1} \\
    \textsuperscript{1}University of Texas at Austin,
    \textsuperscript{2}InterDigital \\
    \texttt{danjacobellis@utexas.edu, neeraja@austin.utexas.edu} \\
    \texttt{\{mateen.ulhaq, fabien.racape, hyomin.choi\}@interdigital.com}
}

\begin{document}
\maketitle

\begin{abstract}
Video comprises the vast majority of bits that are generated daily, and is the primary signal driving current innovations in robotics, remote sensing, and wearable technology.
Yet, the most powerful video understanding models are too expensive for the resource-constrained platforms used in these applications.
One approach is to offload inference to the cloud; this gives access to GPUs capable of processing high-resolution videos in real time.
But even with reliable, high-bandwidth communication channels, the combined latency of video encoding, model inference, and round-trip communication prohibits use for certain real-time applications.
The alternative is to use fully local inference; but this places extreme constraints on computational and power costs, requiring smaller models and lower resolution, leading to degraded accuracy.
To address these challenges, we propose DeDelayed, a real-time inference system that divides computation between a remote model operating on delayed video frames and a local model with access to the current frame.
The remote model is trained to make predictions on anticipated future frames, which the local model incorporates into its prediction for the current frame.
The local and remote models are jointly optimized with an autoencoder that limits the transmission bitrate required by the available downlink communication channel.
We evaluate DeDelayed on the task of real-time streaming video segmentation using the BDD100K driving dataset.
For a round trip delay of 100 ms, DeDelayed improves performance by 6.4 mIoU compared to fully local inference and 9.8 mIoU compared to remote inference---an equivalent improvement to using a model ten times larger.
We release our training code, pretrained models, and python library at \url{https://github.com/InterDigitalInc/dedelayed}.
\end{abstract}
\thispagestyle{firstpage}

\section{Introduction}

\begin{figure*}[t]
  \includegraphics[width=\textwidth]{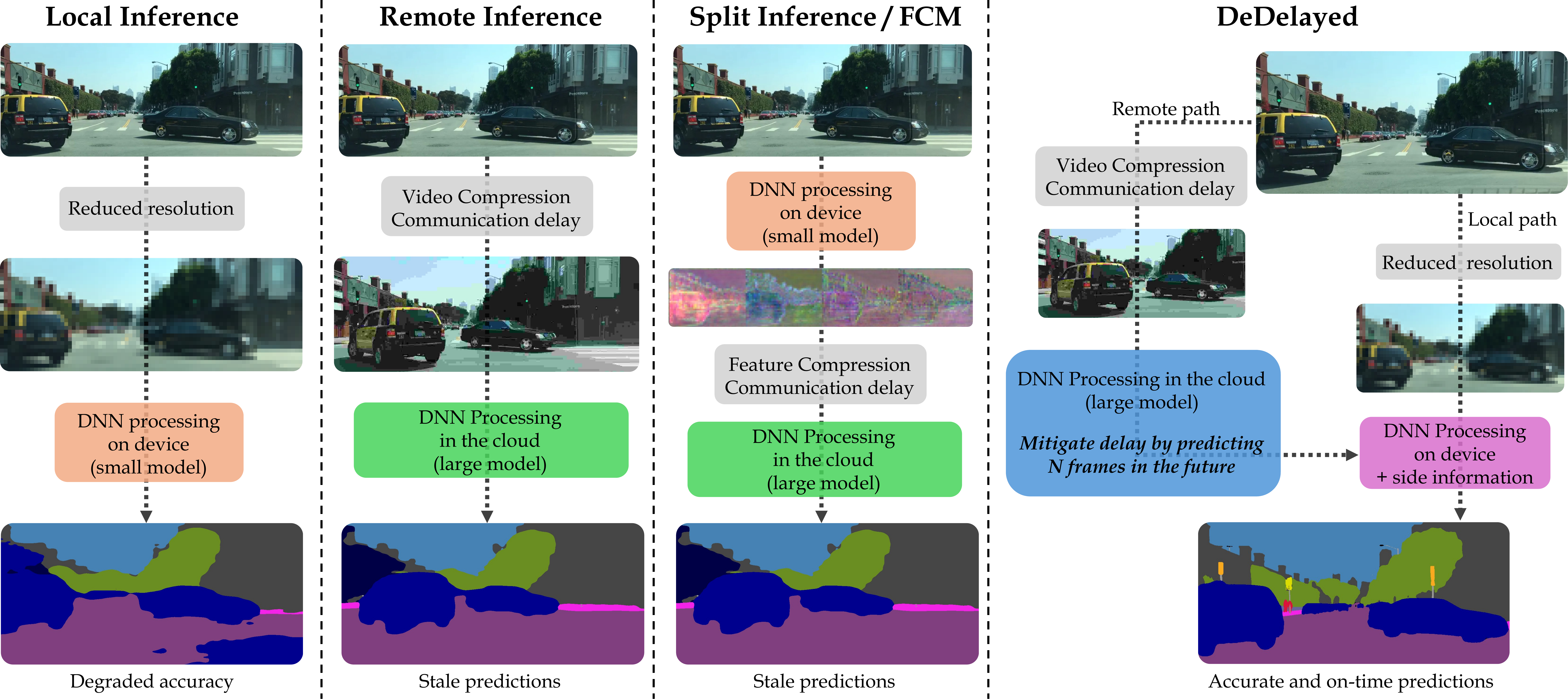}
  \caption{
    Overview of various inference setups, including conventional local, remote, and split inference.
    DeDelayed combines a small on-device image model and a heavier cloud-based temporally predictive video model to produce accurate and on-time predictions.
  }
  \label{fig:dedelayed}
\end{figure*}

In soft real-time applications---such as cloud gaming or video conferencing~\cite{jiang2025real, isikdogan2020eye, jain2025ntire}---late outputs may be diminished in value, but are still useful.
In these applications, expensive Deep Neural Network (DNN)-based operations can be offloaded to powerful cloud GPUs to save on-device power.
As long as the typical latency is low, the loss of utility from latency is outweighed by power savings and extended battery life.

In hard real-time applications---such as aerial robotic control or obstacle avoidance~\cite{liu2022large,moon2021real}---late outputs can be catastrophic, and the system must be designed with a guaranteed deadline.
Due to the irreducible high-tail latency in wireless communication, hard real-time applications must be equipped with a fully functional local inference pipeline as a fallback in the case that the remote predictions fail to meet the deadline~\cite{chen2024fogros2, chen2025fogros2}.

Modern hardware for capturing digital video can operate with extremely low power while ingesting super-human amounts of information at high spatial and temporal resolution~\cite{engel2023project}.
Using video as the primary or exclusive modality of perception provides an opportunity to significantly reduce size, weight, and power by replacing multiple specialized sensors, and is a major focus of ongoing research.
Existing video understanding systems based on masked prediction and diffusion-based foundation models~\cite{ravi2024sam, agarwal2025cosmos, kim2024openvla} are able to directly produce useful predictions or actions from video input, but require too much computation to run directly on the sensor devices used in these applications.
Various approaches to split computing and collaborative inference~\cite{kang2017neurosurgeon, matsubara2022split, wang2024end, gao2025feature} have been proposed to offload computation of these expensive image and video models to the cloud.
For real-time streaming video applications, existing approaches still fall into one or more of three common pitfalls.
(1) They allocate all on-device power and computation to a single linear inference pipeline, leaving no resources for a local-only fallback. \hspace{0pt}
(2) They do not account for the impact of latency on prediction accuracy. \hspace{0pt}
(3) They operate on videos with significantly reduced spatiotemporal resolution to manage computational cost, leaving out rich visual details available from modern camera systems.

\begin{figure}[t]
  \centering
  \includegraphics[width=\linewidth]{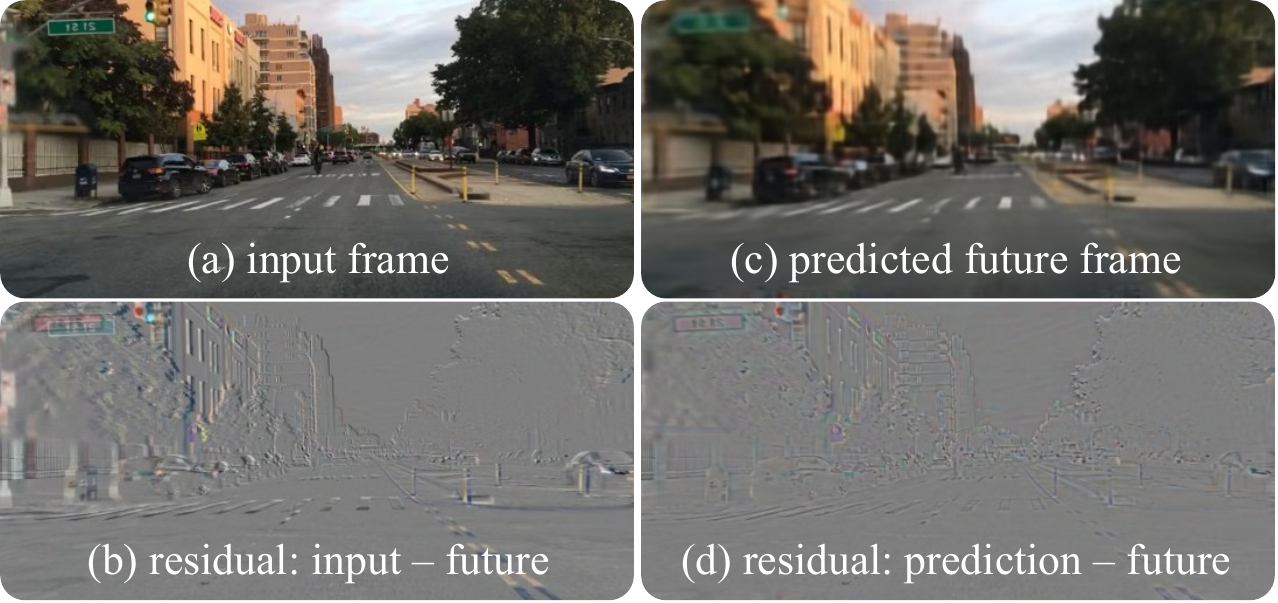}
  \caption{
      To demonstrate the effect of temporally predictive training, we train a 3D transformer to predict the next frame with an MSE loss on pixels.
      (a) shows the original video frame.
      (b) shows the difference between (a) and a future frame, with objects such as the traffic sign and road markings in different locations.
      (c) shows the pixel predictions of the 3D transformer.
      (d) shows the difference from the true future frame.
      While the predictive model cannot predict high-frequency details, it is able to accurately model the motion of objects, signs, and road markings.
  }
  \label{fig:ffp}
\end{figure}

\begin{figure}[t]
  \centering
  \includegraphics[width=\linewidth]{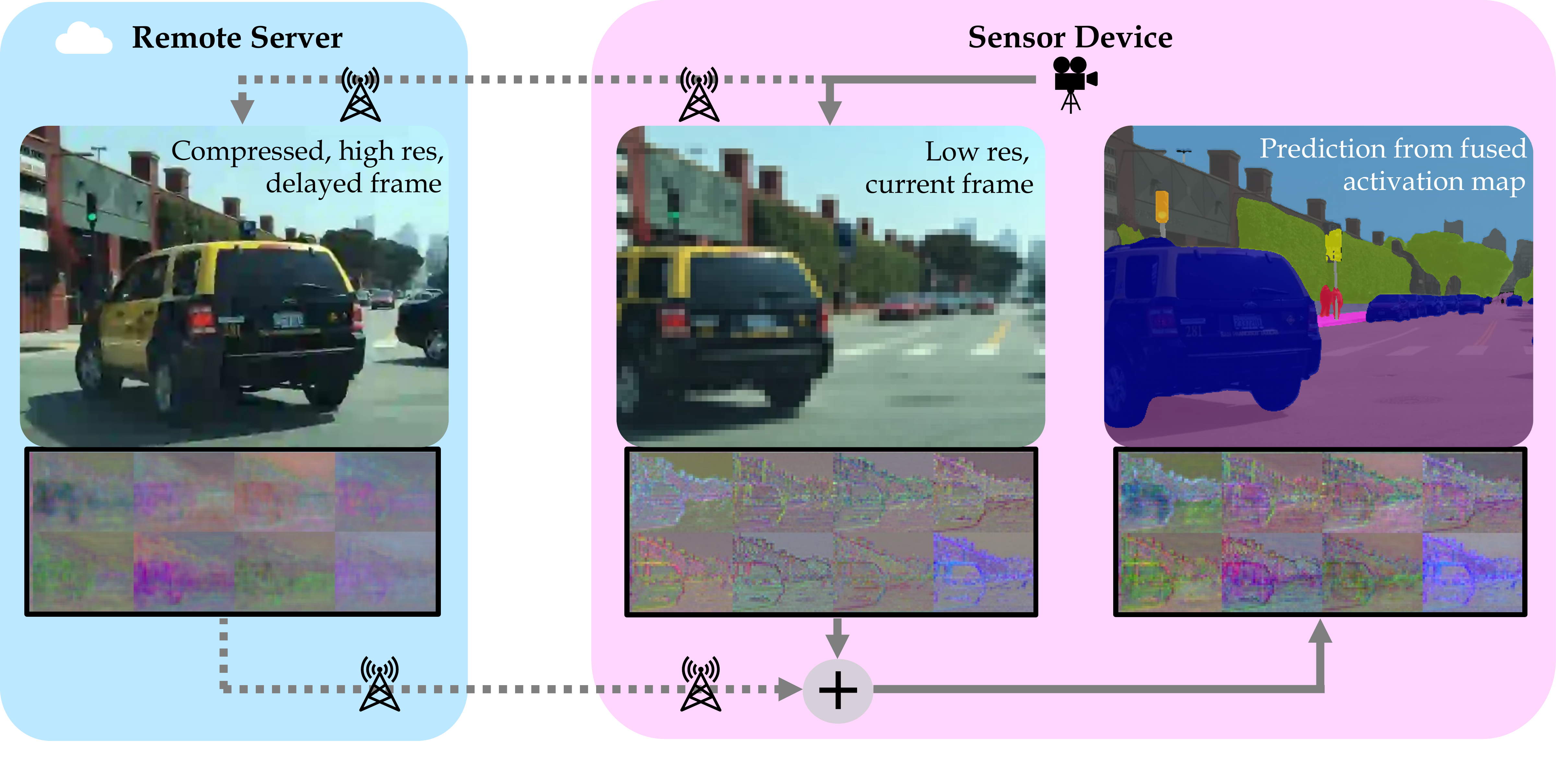}
  \caption{
    Example of activation maps from local and remote model components.
    The remote server uses the higher level of video detail to accurately distinguish and classify objects.
    The local model provides exact position adjustments based on the current frame.
    When making predictions from the combined activation map, small details that would be impossible to make out at low resolution (e.g., the distant pedestrians, labeled red) are accurately classified and localized.
  }
  \label{fig:activation}
\end{figure}

To address these limitations, we introduce DeDelayed~(\cref{fig:dedelayed}), a collaborative inference framework suitable for hard real-time applications.
DeDelayed is designed to maximally utilize high-resolution, full-motion video understanding models on cloud GPUs, while avoiding the pitfalls of other split-computing systems:

\begin{enumerate}
    \item \textbf{Full integration with local-only fallback model.}
    No wireless communication channel can offer perfect reliability.
    For real-time applications with critical deadlines, any remote inference procedure must be accompanied by a lightweight local fallback model.
    Instead of two redundant inference pipelines, DeDelayed uses a single path based on a local model that \emph{optionally} incorporates side information from the remote model.
    We choose a simple method to incorporate this side information---element-wise addition of activation maps---resulting in negligible overhead and well-defined behavior in the absence of remote outputs.
    \hspace{0pt}

    \item \textbf{Temporal prediction for latency mitigation.}
    During supervised training of the remote model component, we simulate a delay of $D$ frames.
    In other words, the remote model is trained to predict features useful in the future.
    A delay embedding---similar to a position embedding in text or vision transformers---allows the behavior of the remote model to adapt to changes in the channel.
    As shown in~\cref{fig:ffp}, temporally predictive training is able to capture motion dynamics, which can be used to compensate for latency.

    \item \textbf{Mixed-resolution inference.}
    On-device AI video processing at capture resolution and frame rate is rarely feasible, even with lightweight models.
    DeDelayed enables mixed-resolution inference---the local model runs at a lower resolution, and the remote model processes high-resolution frames with a 3D transformer that understands motion.
    Thus, the remote model supplies delayed yet accurate, high-level semantic features, while the local model aligns and localizes them to the current scene, as shown in~\cref{fig:activation}.
\end{enumerate}

Our contributions are threefold:

\begin{enumerate}
    \item We provide measurements demonstrating how higher degrees of latency hurt the accuracy of dense visual predictions for semantic segmentation of driving scenes.
    \item We introduce DeDelayed, a co-inference framework that integrates the output of a future-predicting remote model with the current input to a local model.
    \item Using DeDelayed, we create a video segmentation system for urban driving scenes that outperforms any existing local or remote inference solution, while avoiding the pitfalls that limit the practicality of previous approaches.
\end{enumerate}

\FloatBarrier

\section{Background}

Machines equipped with digital video sensors---which are at the center of ongoing innovation in robotics~\citep{kim2024openvla,o2024open}, remote sensing~\citep{szwarcman2024prithvi,khani2021realtimemodelstreaming}, and wearable technology~\citep{grauman2022ego4d,grauman2024ego}---produce high-bandwidth streams that require significant information processing.
The throughput and power efficiency of ingesting pixels on the sensor device (e.g., a battery-powered robot) are extremely high---typically tens or hundreds of megapixels per watt-second~\citep{engel2023project}.
However, moderately sized DNNs can only process visual data at about one megapixel per watt-second~\citep{cai2023efficientvit}.
For more advanced video AI based on autoregressive modeling~\citep{agarwal2025cosmos} or temporal prediction~\citep{assran2025vjepa2}, the efficiency may be as low as 500 pixels per watt-second.
Instead of on-device processing, power constraints can be circumvented by compressing and transmitting video streams to cloud GPU datacenters supported by a 100-megawatt power infrastructure~\citep{goldberg2013cloud}.

Nevertheless, fully remote processing is challenging for certain real-time applications (e.g., collision avoidance) due to unreliability in network and cloud infrastructure~\citep{chen2024fogros2,chen2025fogros2}.
This has motivated split computing systems~\cite{teerapittayanon2016branchynet,kang2017neurosurgeon,mpegAI2025,choi2018deepfeaturecompression,choi2018nearlosslessdeepfeaturecompression,choi2022scalableimagecoding,azizian2022privacy} which can reduce the bandwidth and latency of remote inference by leveraging both on-device and remote computation.
Still, delivering predictions by a guaranteed deadline requires a fallback procedure independent of the remote server.
In many systems (e.g., autonomous motor vehicles) the limited accuracy and reliability of lightweight local models warrant a human operator as the fallback~\citep{on2021taxonomy}, preventing full automation.

\section{Related work}

The challenges of real-time video perception have been approached from many angles.
Previous research can be grouped into four major categories:
(1) improving the accuracy-efficiency trade-off of inference,
(2) offloading computation from the sensor device to a remote server,
(3) temporally predictive video modeling, and
(4) dynamically augmenting the on-device model via remote collaboration.

\paragraph{Improving the accuracy-efficiency trade-off.}

Over the decade since AlexNet~\cite{krizhevsky2012alexnet}, the trade-off between accuracy and computational efficiency of image-based DNNs has improved significantly.
This has made local inference (shown in the first column of~\cref{fig:dedelayed}) the de facto standard for many computer vision and robotics applications, as the target perception accuracy is often achievable on-device without introducing the additional failure modes or network requirements.
Still, video models that process dense 3D pixel volumes at high resolution using convolutional or transformer-based DNNs are typically avoided for real-time applications.
Instead, it is common to apply a lightweight 2D image model to each frame.
Using recent 2D models like EfficientViT~\cite{cai2023efficientvit}, high resolution images can be segmented accurately at real-time throughput (greater than 30 fps) using moderately powered hardware (10--100 Watts).
For more restricted power budgets (less than 10~Watts) the spatial or temporal resolution must be reduced by one or more orders of magnitude to meet real-time deadlines, significantly reducing accuracy.

\paragraph{Offloading computation.}

Numerous works have explored ways to offload computation of expensive image and video models to the cloud.
Early exiting~\cite{teerapittayanon2016branchynet}, split computing~\cite{kang2017neurosurgeon}, and feature coding for machines (FCM)~\cite{mpegAI2025,choi2018deepfeaturecompression,choi2018nearlosslessdeepfeaturecompression,choi2022scalableimagecoding,azizian2022privacy} focus on partitioning DNN layers into two components: one that runs on the sensor device and another that runs in the cloud.
The overall workflow of these approaches is shown in the third column of~\cref{fig:dedelayed}.
A limitation of these methods and many of their successors is that they allocate local computation towards a prediction pipeline that terminates in the cloud, rather than on the device.
This may be desirable for applications where the predictions are only used in the cloud (e.g., remote sensing).
However, if predicted outputs must be available on-device (e.g., autonomous navigation), the delay incurred by round-trip communication leads to stale predictions.
Even worse, the exclusive allocation of local computational resources to an inference pipeline that necessarily requires a remote server leaves little room for a readily available local fallback mode in the event of late or dropped predicted outputs.

\paragraph{Temporally predictive video modeling.}

Regardless of DNN size or added network latency, some amount of inference delay is unavoidable.
Instead of focusing exclusively on reducing the degree of delay, prior works have explored correcting the estimated world state based on the anticipated delay.
These corrections require some mechanism for temporally predictive video modeling, and can use on-device optical flow estimates~\cite{zhang2024neuflow} or readily available motion vectors from an existing video codec~\cite{xiao2024task}.
If video is streamed to a sufficiently powerful remote server, generative video world models based on masked prediction~\cite{assran2025vjepa2} or diffusion~\cite{agarwal2025cosmos} foundation models can provide temporal prediction.
In each of these cases, existing methods for temporal prediction are repurposed from another application, rather than specialized and optimized for the desired perception task.
This prevents temporally predictive abilities or behaviors from being learned or optimized with the rest of the DNN-based layers.

\paragraph{Dynamically augmenting the on-device model via remote collaboration during streaming inference.}

This category of methods improve the on-device model's inference capabilities by incorporating remote information at runtime (e.g., streaming weights or fusing remote features).
They are underexplored but important for real-time applications on computationally limited devices.
Adaptive Model Streaming~\cite{khani2021realtimemodelstreaming} streams weight updates to keep the local model tuned to changing scene characteristics.
Knowledge Boosting~\cite{srinivas2024knowledgeboosting} addresses the problem from a different lens---instead of dynamically updating the model weights, it directly fuses in remote features into a small on-device model.
It requires separate models trained for specific fixed delays, and evaluates on audio data in a controlled local environment with delays up to 48~ms within a local networking context.
Our method, DeDelayed, is capable of adapting to arbitrary remote inference delays by conditioning the remote model on the measured delay at runtime, enabling dynamic specialization across varying latency.
Methods within this category are orthogonal to and compatible with many methods from earlier categories, and with each other.
Our work extends this rarely-explored direction by explicitly inputting dynamic system characteristics such as delay to improve on-device predictions under realistic, variable latency.

\section{Method}

\begin{figure}[tbp]
    \centering
    \includegraphics[width=1\linewidth]{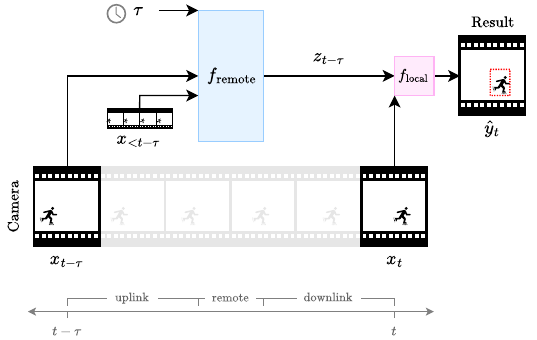}
    \caption{
        Time progresses left to right.
        The client-side camera produces video frames, which are sent across a communication network to the server.
        The server runs a heavyweight model using the latest video frame $x_{t - \tau}$ that it receives, in addition to a context of previously received video frames $x_{< t - \tau}$, as well as the measured delay $\tau$.
        This produces an output $z_{t - \tau}$ that the server sends to the client.
        The client pairs the latest received response $z_{t - \tau}$ with a freshly produced video frame $x_t$, and runs these inputs through a lightweight model.
        This finally produces a timely result $\hat{y}_t$ that can be used in real-time delay-sensitive applications.
    }
    \label{fig:system_overview}
\end{figure}

DeDelayed introduces a general framework that improves the accuracy and robustness of real-time inference on resource-constrained sensor devices.
It does so by combining the strengths of both local inference and remote inference, while mitigating their weaknesses.
The local model has access to the latest sensor data, and yet lacks the computational capability needed to produce accurate outputs.
The remote model provides accurate outputs, and yet delivers them with delay.
With careful combination of both subsystems, DeDelayed is able to provide bounded performance guarantees.
When trained appropriately such that the local and remote subsystems deliver the maximum accuracies that they would conventionally---e.g., by freezing the local and remote image backbones during the final joint training of the entire system---it can be verified that DeDelayed is never worse than either local inference or remote inference independently.
As we will demonstrate later, we are able to glue together the two subsystems in a way that is simple yet effective.

DeDelayed addresses the problem of stale predictions from powerful remote models by integrating them with a lightweight, on-device model.
The core idea is to leverage the high-quality features from a heavyweight remote model, despite their inherent delay, by explicitly conditioning them on the measured latency and fusing them early with live information from a local model.
This ensures that the final predictions are both accurate and timely.

DeDelayed can be formulated in simple mathematical terms as follows.
Given a fresh input frame $x_t$ at current time $t$, the final prediction $\hat{y}_t$ is computed using a lightweight local model, $f_{\text{local}}$, which processes $x_t$ along with time-delayed features $z_{t-\tau}$ from a heavyweight remote model, $f_{\text{remote}}$.
To produce powerful predictive features, the remote model is conditioned on the delay $\tau$, and processes a short clip of past frames $x_{\leq t-\tau}$ ending at time $t-\tau$.
This is expressed by the following equations:
\begin{gather}
z_{t - \tau} = f_{\text{remote}}(\tau,\, x_{\leq t - \tau}) \label{eq:remote_features} \\
\hat{y}_t = f_\text{local}(x_t,\, z_{t - \tau}) \label{eq:final_prediction}
\end{gather}
For clarity, the notation is summarized in~\cref{tbl:symbol-meaning}.
\cref{fig:system_overview} presents a system diagram that demonstrates the fundamental principle we describe, and shows how information propagates through the various subsystems as time progresses.

\begin{table}[h]
\caption{Notation.}
\label{tbl:symbol-meaning}
\begin{center}
\small
\setlength{\tabcolsep}{3pt}
\begin{tabular}{ll}
\textbf{Symbol} & \textbf{Meaning} \\
\hline
\addlinespace[2pt]
$x_t$ & Input frame at current time $t$ \\
$x_{\leq t - \tau}$ & Input frames up to time $t - \tau$ \\
$\hat{y}_t$ & Prediction for time $t$ \\
$z_{t - \tau}$ & Features outputted by remote model run at time $t - \tau$ \\
$\tau$ & Delay in time between the old and current frame \\ 
$D$ & Delay in frames between the old and current frame \\ 
$f_\text{local}$ & Light local model run at time $t$ \\
$f_\text{remote}$ & Heavy remote model run at time $t - \tau$ \\
\end{tabular}
\end{center}
\end{table}

The entire DeDelayed system is trained end-to-end to minimize a task-specific loss function, $\mathcal{L}_{\text{task}}$, evaluated against the ground truth $y_t$ for the current frame.
\begin{equation*}
    \mathcal{L}_{\text{task}} = \ell(\hat{y}_t, y_t)
\end{equation*}
For semantic segmentation, $\ell$ is typically the cross-entropy loss.
The objective is to produce predictions $\hat{y}_t$ that are accurate at time $t$ on the local device.

In the next section, we detail how we designed a specific implementation to test the DeDelayed framework in action.

\begin{figure*}[h]
  \centering
  \includegraphics[width=1\linewidth]{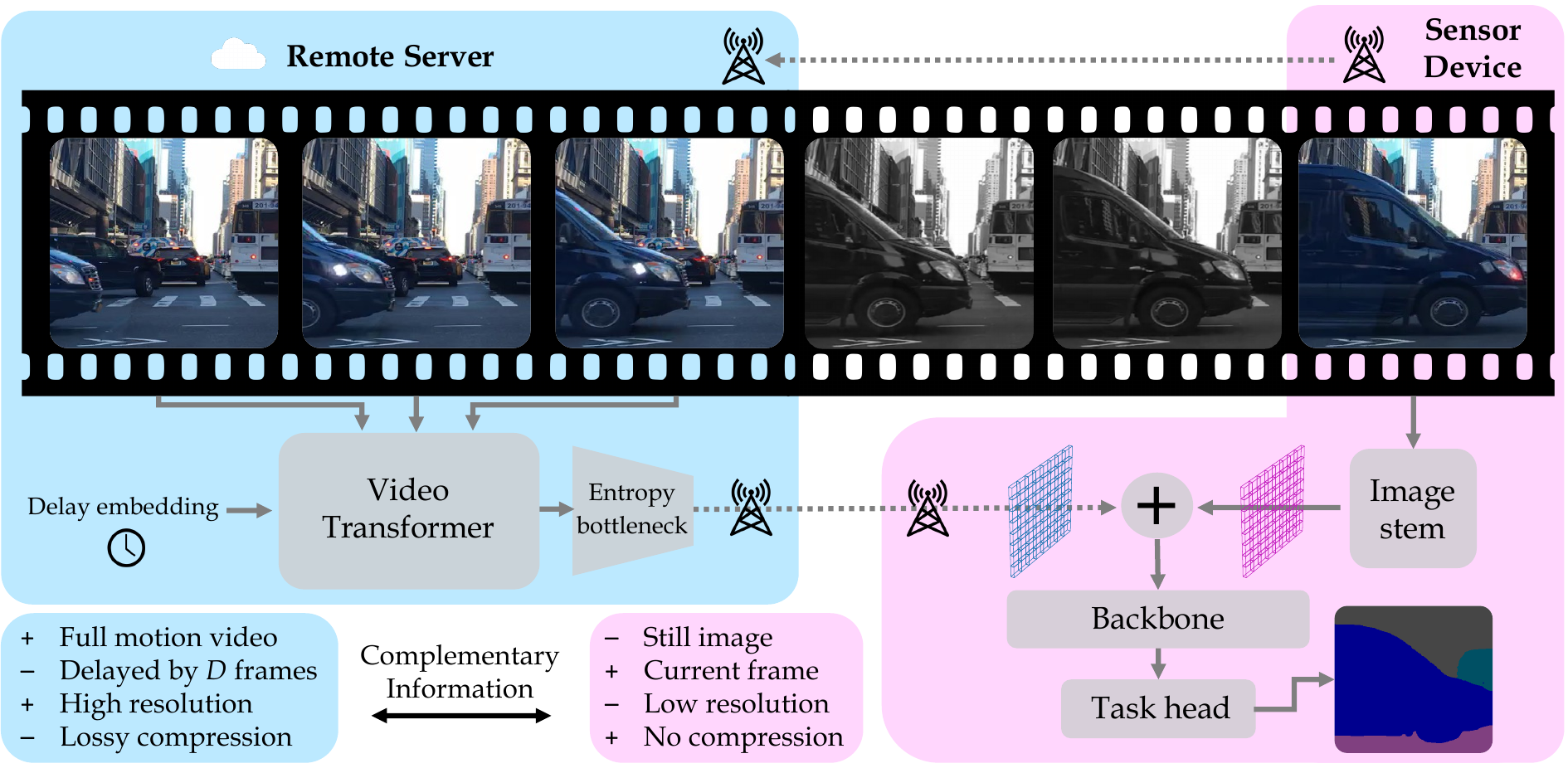}
  \caption{
    DeDelayed workflow.
    The complementary properties of local and remote inference are exploited to produce real-time streaming video understanding that is both accurate and on time.
  }
  \label{fig:design}
\end{figure*}

\section{Design and implementation}

DeDelayed aims to use remote computation to improve the accuracy and robustness of real-time machine vision on resource constrained sensor devices.
The design of DeDelayed is guided by three key insights.
(1) Incorporating the outputs of the remote model as side information to the local model eliminates computational redundancy and provides a fallback suitable for hard real-time applications.
(2) Temporally predictive training can mitigate the impact of latency due to processing and round-trip communication.
(3) Decoupling inference into a high resolution full-motion video stream and a low-resolution still-image stream exploits the complementary properties of remote and on-device processing.
\cref{fig:design} shows an overview of the design in the context of semantic segmentation of video frames.

\paragraph{System overview.}

Information propagates through our system as follows:

\begin{enumerate}
  \item The local device transmits input frames to the remote via the uplink.
  \item Each incoming frame is fed into a heavyweight remote model $f_{\text{remote}}$.
  To accelerate inference, this model caches and maintains a context window of the $K$ most recent features, which are computed from each incoming frame using a pretrained 2D ViT backbone.
  \item The $K$ per-frame features are concatenated along the temporal axis, and a learned delay embedding conditioned on the measured delay $\tau$ is added.
  \item A 3D ViT encoder followed by learned pooling (MLP–pool–MLP) produces delay-conditioned remote features $z_{t-\tau}$, which are sent back to the device via the downlink.
  \item The lightweight local model $f_{\text{local}}$ runs on a fresh input $x_t$, and fuses in the remote features $z_{t-\tau}$.
  \item The local model finishes decoding the fused representation and outputs labels $\hat{y}_t$.
\end{enumerate}

\paragraph{Video compression.}

Each frame captured on the sensor device is processed by a lossy image or video codec to allow transmission over a wireless channel.
In our experiments, we choose the resolution, framerate, and degree of lossy compression to represent a video signal that is transmissible via 5G cellular uplink (30fps 720p content compressed at rates between 1--10 Mbps).

\paragraph{Latency-aware remote video model.}

\begin{figure*}[t]
  \centering
  \includegraphics[width=1\linewidth]{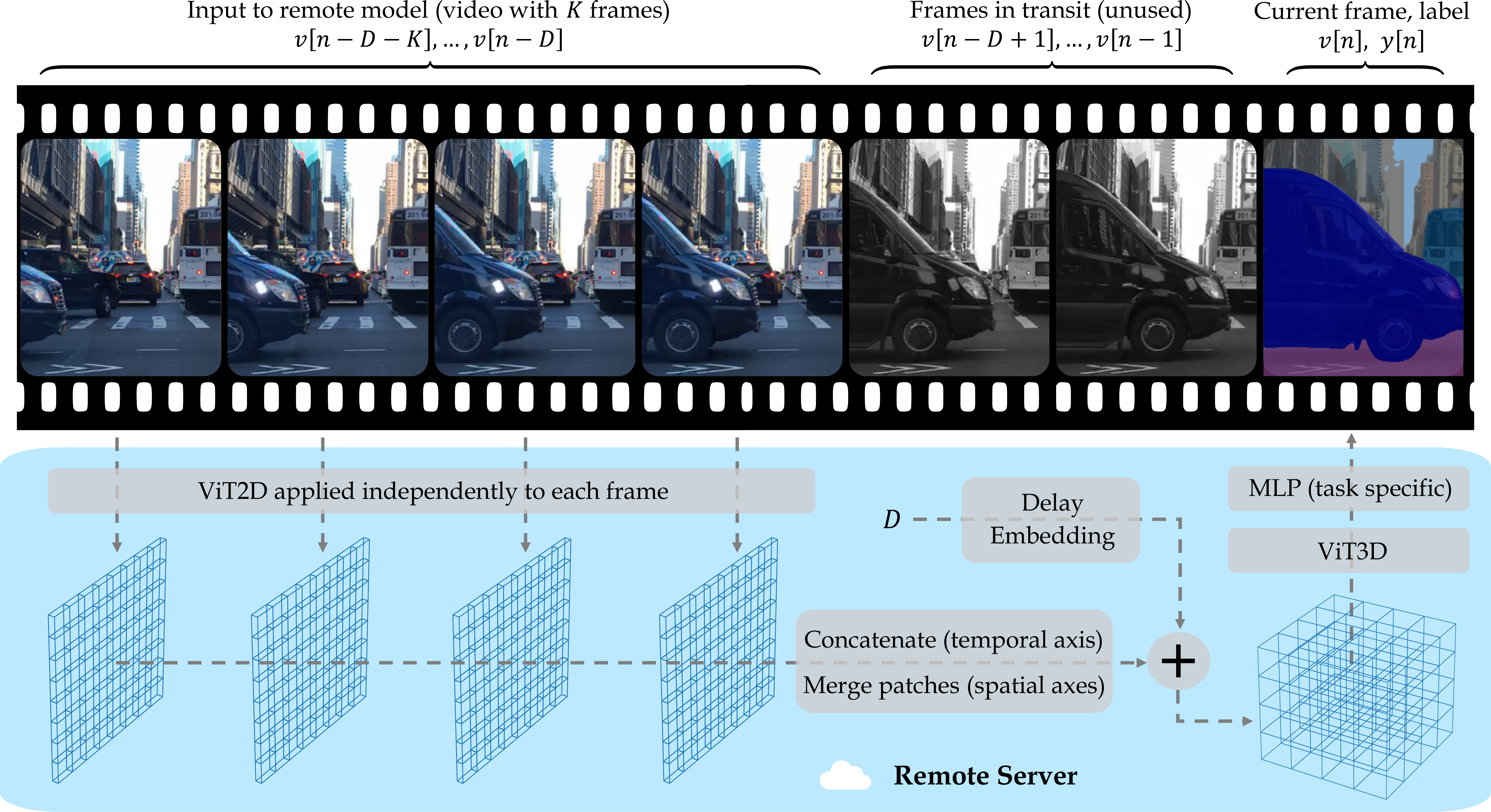}
  \caption{
    Overview of the remote model component as used during pretraining.
    A video sequence $v[n]$ is captured on the sensor device and transmitted to a remote server, incurring a delay of $D$ frames.
    The remote model operates on a fixed context window of $K$ frames.
    A 2D vision transformer (EfficientViT-L1, effective patch size of $8\times8$ pixels) is applied independently to each of the $K$ input frames.
    The outputs of the 2D transformer are concatenated along the temporal axis, but spatially merged into larger $16\times16$ patches, thus maintaining a similar sequence length (when $K=4$, the sequence length is identical).
    A learned delay embedding is added, allowing the remaining layers to behave differently based on the expected value of $D$.
    Finally, 3D video transformer layers and task-specific (e.g., segmentation) layers are applied, generating a prediction corresponding to the current frame ($D$ frames in the future after the latest remote input frame).
  }
  \label{fig:vit3d}
\end{figure*}

The remote model operates on high resolution, compressed video frames, and consists of four learnable modules:
(1) a 2D vision transformer (ViT2D),
(2) a 3D video transformer (ViT3D),
(3) task-specific MLP layers (R-MLP), and
(4) a dimensionality-reducing autoencoder (DR-AE).
During the initial training stages, only modules 1--3 (ViT2D, ViT3D, and R-MLP) are used, as shown in~\cref{fig:vit3d}.
During the final training stage, the R-MLP module is discarded and replaced with the DR-AE to allow joint training of the local and remote components.
Due to round trip communication and networking, the predictions from the remote model will be delayed by $D$~frames at runtime.
During training, an artificial delay is applied to the input, but not to the target.
Thus, the training objective is to predict the target of a future frame.
During training, we sample a delay uniformly between 0 and 5 frames (up to 167~ms).
Additionally, we add a learnable "delay embedding" to the input activation maps of the ViT3D---similar to the position or timestep embedding used in diffusion transformers~\citep{peebles2023scalable}.
In our experiments, we provide the remote model with four frames of context and aim for a target latency of 33~ms (a single frame at 30~fps) on a high power GPU testbed.

\paragraph{Local image model incorporating remote side information.}

The local model processes the most recently collected video frame on the sensor device at low resolution, and consists of three learnable modules, shown in the right half of~\cref{fig:design}:
(1) pixel-processing 2D convolutional layers (CNN2D),
(2) 2D feature-processing convolution and attention layers (CoAt2D), and task specific (e.g., classification or segmentation) MLP layers (L-MLP).
If present, features from the remote model are added element-wise to the activation map between the CNN2D and the CoAt2D.
Our local model can operate at a resolution up to 0.34 megapixels ($704 \times 480$) while meeting the same target latency of 33~ms on our CPU (rather than GPU) testbed.

\paragraph{Joint prediction with autoencoded resolution adapter.}
After pretraining the remote model for the desired prediction task, its task-specific layers are discarded and replaced with the DR-AE, consisting of adaptive spatial pooling and a channel bottleneck.
The degree of adaptive spatial pooling is chosen to match the operating resolution of the local model.
The sample mean is used as the pooling operator.
The remainder of the DR-AE consists of MLP layers that terminate in a low-resolution, low-channel count activation map suitable for downlink transmission.
The output of the DR-AE is added element-wise to the output of the pixel-processing CNN2D.
Then, the local and remote models undergo a final joint training stage.

\section{Evaluation}

\subsection*{Experimental setup}
We evaluate DeDelayed on the task of real-time semantic segmentation of driving scenes using the BDD100K video dataset~\citep{dataset_bdd100k}, containing video of driving scenes at 30 frames per second (fps).
We use the standard 19 label Cityscapes taxonomy~\cite{dataset_Cordts2016Cityscapes} for semantic segmentation of urban scenes.
Since the BDD100K dataset does not provide dense segmentation labels for all video frames, we generate pseudo-labels using two models.
For the validation set, we use the pretrained DepthAnything Cityscapes segmentation model~\cite{yang2024depth_anything_v1}.
To our knowledge, this is the most accurate publicly available semantic segmentation model for the Cityscapes taxonomy.
For the training set, we use EoMT~\citep{kerssies2025eomtvitsecretlyimagesegmentation}, which provides high accuracy, but is significantly faster for labeling 70k training images.

We evaluate the performance subject to known delays ranging from 0 to 5 frames, corresponding to 0 to 167\,ms at 30 fps.
At training time, the delay~$\tau$ is sampled per batch from a uniform distribution over this range.

\subsection*{Training details}

We adopt a multi-stage training strategy, as detailed in~\cref{tbl:training_stages}.
The remote and local models are first trained individually and then later combined.
Each model is pretrained on the large-scale ImageNet dataset~\citep{dataset_ILSVRC15} for classification, then on the image segmentation task on Cityscapes~\citep{dataset_Cordts2016Cityscapes}, before being fine-tuned on the smaller BDD100K driving dataset.
We train the remote model to have temporally predictive capability by supplying it with a delay-aware (DA) objective: to predict the labels of future frames conditioned on the degree of delay.
During the final training stage, the task-specific layers of the MLP are replaced with an entropy bottleneck and resolution adapter that allow its outputs to be added to the intermediate activation map of the local model. Using this configuration, the local and remote models are trained jointly for the target scenario of local segmentation incorporating delayed predictions from the remote model.
The training loss is per-pixel cross-entropy loss. We use the Adan~\citep{xie2024adan} optimizer, a warmup-stable-decay learning rate schedule, gradient clipping, and selectively applying discriminative fine-tuning or layer-wise learning rate decay (LLRD)~\citep{howard2018ulmfit_discriminative_finetuning_LLRD}.

\begin{table}[ht]
    \centering
    \caption{Local and remote model components and training setup.}
    \small
    \begin{tabularx}{\linewidth}{@{}c p{2.3cm} p{2.1cm} X c@{}}
        \toprule
        Stage & Local layers & Remote layers & Data & Res. \\
        \midrule[\heavyrulewidth]
        \multicolumn{5}{c}{\textbf{Remote (video-predictive)}} \\
        \midrule
        \addlinespace[2pt]
        1 & -- & ViT2D & IN1K, CS & -- \\
        2 & -- & ViT2D, ViT3D & BDD & 496 \\
        \midrule[\heavyrulewidth]
        \multicolumn{5}{c}{\textbf{Local (image only)}} \\
        \midrule
        \addlinespace[2pt]
        3 & CNN2D, CoAt2D & -- & IN1K & 224 \\
        4 & CNN2D, CoAt2D & -- & CS & 336 \\
        5 & CNN2D, CoAt2D & -- & BDD & 496 \\
        \midrule[\heavyrulewidth]
        \multicolumn{5}{c}{\textbf{DeDelayed}} \\
        \midrule
        \addlinespace[2pt]
        6 & CNN2D, CoAt2D & ViT2D, ViT3D & BDD & 480/720 \\
        \bottomrule
    \end{tabularx}

    \vspace{2pt}
    \footnotesize

    Data: IN1K = ImageNet-1K; CS = Cityscapes; BDD = Berkeley DeepDrive 100K.

    \label{tbl:training_stages}
\end{table}

\section{Results}

We compare how various inference systems perform under the effect of communication network latency.
\cref{tbl:accuracy_under_delay} shows the segmentation performance (mIoU) for different configurations (local-only, remote-only, and the proposed local+remote system, DeDelayed).
Each configuration serves as an ablation on the final design.
\begin{itemize}
    \item \textbf{Local image} and \textbf{Remote image} inference setups process individual frames in the conventional way, though the remote is susceptible to communication network delay.
    \item \textbf{Remote video} has access to past frames of context, but only predicts labels for its present view, and thus fares no better than "remote image".
    \item \textbf{Remote predictive} is fed a tunable delay and sustains accuracy by predicting the future.
    \item \textbf{Local + remote predictive} represents a DeDelayed system, and thus, is able to further sustain accuracy by merging the remote predictive features with fresh local features.
\end{itemize}

\begin{figure}[htbp]
    \centering
    \includegraphics[width=1\linewidth]{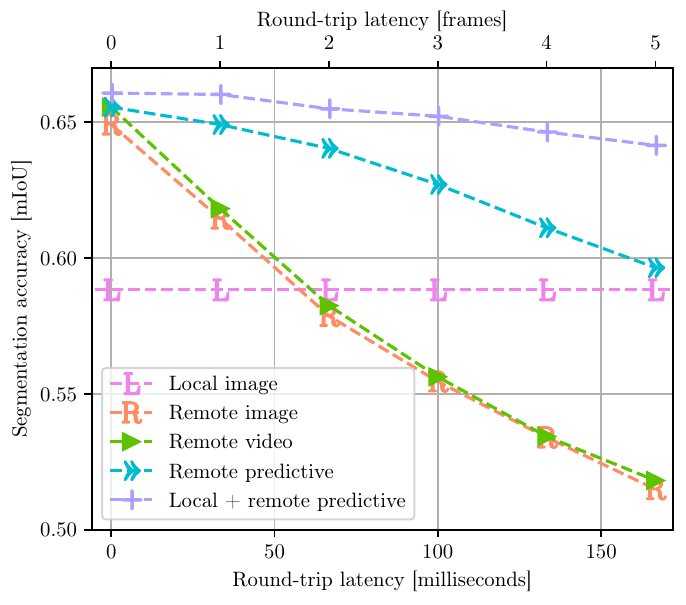}
    \caption{
        Segmentation accuracy (mIoU) versus round-trip latency (milliseconds or frames).
    }
    \label{fig:miou_latency}
\end{figure}

As illustrated in~\cref{fig:miou_latency}, the performance of conventional remote inference is significantly higher (roughly 5\%pt.~mIoU) than local inference if communication delay is low.
However, the accuracy degrades rapidly with increasing network delay, reflecting the detriment of using stale predictions for dense prediction tasks involving motion.
Beyond 2 frames (67~ms) of round-trip latency, remote inference becomes worse than local inference (8\%pt.~mIoU lower at 167~ms).
This impact is significantly mitigated by adopting a temporal prediction training objective, which extends the utility of remote inference to 5 frames (167~ms).
Finally, the proposed configuration (DeDelayed) incorporates temporally predictive features from the remote model as side information to the local model, and almost completely eliminates the drop-off in accuracy due to delay.
At the maximum tested delay of 5 frames, DeDelayed provides a 6.7\%pt.~mIoU improvement over the local baseline---a difference that would typically require using a model roughly 10 times larger~\cite{cai2023efficientvit, kerssies2025eomtvitsecretlyimagesegmentation}.

\cref{tbl:accuracy_under_delay} shows the accuracy of various inference systems under various local and remote subsystem delays.
For local inference delays of $\leq$8\,ms, the "local + remote predictive" method consistently delivers better accuracy across all network round-trip delays.

\begin{table}[ht]
    \centering
    \caption{Accuracy (mIoU) given various subsystem delays (ms).}
    \footnotesize
    \begin{tabularx}{\linewidth}{@{}c | c c c c c c@{}}
        \toprule
        Local delay & \multicolumn{6}{c}{Remote delay} \\
        & 0 ms & 33 ms & 67 ms & 100 ms & 133 ms & 167 ms \\
        \midrule[\heavyrulewidth]
        \multicolumn{7}{c}{\textbf{Remote (video-predictive)}} \\
        \midrule
        \addlinespace[2pt]
        -- & 0.655 & 0.649 & 0.640 & 0.627 & 0.611 & 0.596 \\
        \midrule[\heavyrulewidth]
        \multicolumn{7}{c}{\textbf{Local (image only)}} \\
        \midrule
        \addlinespace[2pt]
         0 ms & 0.588 & 0.588 & 0.588 & 0.588 & 0.588 & 0.588 \\
         4 ms & 0.585 & 0.585 & 0.585 & 0.585 & 0.585 & 0.585 \\
         8 ms & 0.582 & 0.582 & 0.582 & 0.582 & 0.582 & 0.582 \\
        33 ms & 0.562 & 0.562 & 0.562 & 0.562 & 0.562 & 0.562 \\
        \midrule[\heavyrulewidth]
        \multicolumn{7}{c}{\textbf{DeDelayed}} \\
        \midrule
        \addlinespace[2pt]
         0 ms & 0.661 & 0.660 & 0.655 & 0.652 & 0.646 & 0.641 \\
         4 ms & 0.656 & 0.655 & 0.650 & 0.647 & 0.642 & 0.637 \\
         8 ms & 0.652 & 0.650 & 0.646 & 0.643 & 0.637 & 0.632 \\
        33 ms & 0.624 & 0.619 & 0.616 & 0.612 & 0.607 & 0.603 \\
        \bottomrule
    \end{tabularx}

    \vspace{2pt}
    \footnotesize

    Sub-frame (4~ms, 8~ms) accuracies were lerped between 0~ms and 33~ms.

    \label{tbl:accuracy_under_delay}
\end{table}

\paragraph{Additional experiments.}
\label{sec:results/remote_jitter}
We also investigate the impact of remote delay jitter on system accuracy.
Since the local model is stateless, the only thing important to its accuracy is what delay the received remote features were conditioned on.
In general, the most recently received remote features provide the largest boost to local model accuracy.
If the remote features are extremely stale, or are missing, the local model may simply ignore them during inference.
The supplementary material contains additional ablations, including a delay-input ablation, robustness to uplink compression, a compute breakdown, and evaluation on the Nymeria dataset~\cite{dataset_nymeria}.

\section{Conclusion}

DeDelayed addresses a central challenge in real-time systems that rely on remote computation: prediction staleness induced by network delay.
It mitigates remote inference delay by elevating delay to a first-class variable, conditioning the remote model via a learnable delay embedding, and fusing remote features with fresh local features.
Across realistic network conditions, DeDelayed surpasses strong local-only and remote-only baselines, with a particular advantage for longer latencies and high-motion content.
As a foundational framework, DeDelayed applies to a wide range of real-time problem domains, enabling intelligent systems that are not only accurate but also truly timely and dependable in dynamic environments.
Future work includes studying variable and stochastic delay distributions, high-motion data, lighter local models, and local future prediction.

{
    \small
    \bibliographystyle{ieeenat_fullname}
    \bibliography{main}
}

\clearpage
\twocolumn[
  \centering
  \Large
  \textbf{Appendix}\\
  \vspace{1.0em}
]

\subsection{Code}

The fused "local + remote predictive" model is defined as:

\begin{lstlisting}[
    % float=*t,
    basicstyle=\ttfamily\tiny,
    language=Python,
    mathescape=true,
]
class FusedModel(nn.Module):
    def __init__(self):
        super().__init__()
        self.local_model = LocalModel()
        self.remote_model = RemoteModel()

    def forward(self, x_local, x_remote, delay):
        # Ensure remote z size matches local h size:
        H_local, W_local = x_local.shape[-2:]
        z_size = (H_local // 8, W_local // 8)
        # Receive remote features:
        z = self.remote_model(x_remote, delay, z_size, head="features")
        # Local model with optional fusion:
        y = self.local_model(x_local, z)
        return y


class RemoteModel(nn.Module):
    def __init__(self, name="efficientvit-seg-l1-cityscapes"):
        super().__init__()
        self.image_model = create_efficientvit_seg_model(name)
        self.video_model_backbone = nn.Sequential(*[
            VitBlock3D(in_channels=256, head_dim=32, expand_ratio=4)
            for _ in range(12)
        ])
        self.delay_embedding = DelayEmbedding(output_shape=(256, 1, 1, 1))
        self.seg_head = RemotePredictiveHead(seg_classes=19)
        self.mlp_pre_pool = PrepoolBlock()
        self.mlp_post_pool = PostpoolBlock()

    def forward(self, x_remote, delay, z_size, head):
        # Apply 2D image model backbone to each frame individually:
        z = x_remote
        z = einops.rearrange(z, "b c f h w -> (b f) c h w", f=4)
        z = self.image_model.backbone(z)
        z = einops.rearrange(z, "(b f) c h w -> b c f h w", f=4)
        # Add delay embedding element-wise, then run 3D ViT:
        z = z + self.delay_embedding(delay)
        z = self.video_model_backbone(z)
        # Collapse temporal axis into channel axis, and apply 2D head:
        z = einops.rearrange(z, "b c f h w -> b (c f) h w", f=4)
        if head == "logits":  # Send labels (no fusion).
            z = self.seg_head(z)
        elif head == "features":  # Send features (for fusion).
            z = self.mlp_pre_pool(z)
            z = F.adaptive_avg_pool2d(z, z_size)
            z = self.mlp_post_pool(z)
        return z


class LocalModel(nn.Module):  # Any 2D image segmentation model.
    def forward(self, x_local, z_remote=0):
        # Local feature extraction:
        h = self.T1(x_local)
        # Local and remote fusion:
        h = h + z_remote
        # Remaining local model:
        y = h
        h = self.T2(h)
        y = y + self.P2(h)
        h = self.T3(h)
        y = y + self.P3(h)
        y = self.seg_head(y)
        return y
\end{lstlisting}

During training and evaluation, the models are fed frames and target labels offset by the appropriate delay.
Otherwise, training minimizes per-pixel cross-entropy loss and evaluation uses mIoU as usual.

\FloatBarrier

\subsection{Additional architecture details}

\paragraph{Delay embedding.}
We use a two-layer MLP which maps $\mathbb{R} \to \mathbb{R}^C$ via layers of sizes 1024 and 256.

\paragraph{ViT3D.}
This consists of 12 blocks with 256 input/output channels, where each block contains a 3D attention followed by $3 \times 3 \times 3$ MBConv3D.

\FloatBarrier

\subsection{Effect of delay jitter}

\begin{figure}[htbp]
    \begin{center}
    \includegraphics[width=1\linewidth]{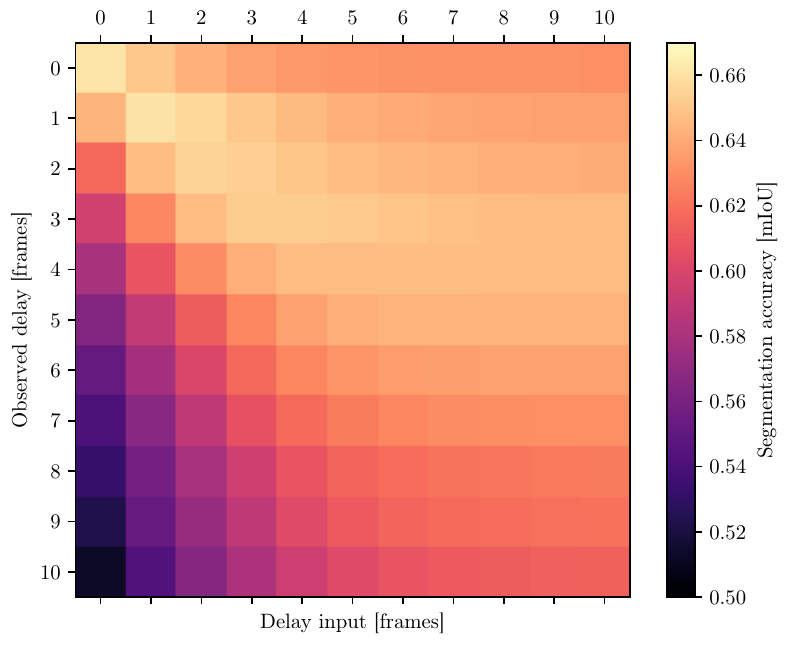}
    \end{center}
    \caption{
        Segmentation accuracy (mIoU) over observed delay and model delay input.
    }
    \label{fig:miou_by_delay_matrix}
\end{figure}

We evaluate how our model performs under delay jitter, i.e., when the delay varies over time.
Our training loss targets accuracy for a fixed, tunable delay input---we do not explicitly train it to be jitter-resilient.
Nonetheless, temporal structure in the data helps maintain accuracy even when the delay input differs from the observed delay.
\cref{fig:miou_by_delay_matrix} characterizes this, showing performance across observed delays when the model is fed a possibly incorrect delay as input.

Although the model was not explicitly trained for mismatched delays or delays beyond 5 frames, it still performs well at these out-of-distribution delays.
Unsurprisingly, accuracy peaks when the model's delay input matches the observed delay.
Notably, the accuracy drop is smaller when the delay input exceeds the observed delay than the reverse.
Thus, when jitter is high, it is safer to use a larger delay input than expected, since underestimating the delay tends to make the remote features overconfident about localization.

This matrix can be precomputed during evaluation.
As noted in \cref{sec:results/remote_jitter}, at runtime the device may consult this matrix to choose among the received remote feature tensors $\{z_1, z_2, \ldots\}$ by looking up the accuracy for each tensor's delay pair $(D_o, D_i)$ and selecting the best.
After a dramatic scene change or excessive staleness, the device may omit remote features from the local model input and run the local model alone, yielding its baseline performance.


\FloatBarrier

\subsection{Local input resolution}

\begin{figure}[htbp]
    \begin{center}
    \includegraphics[width=1\linewidth]{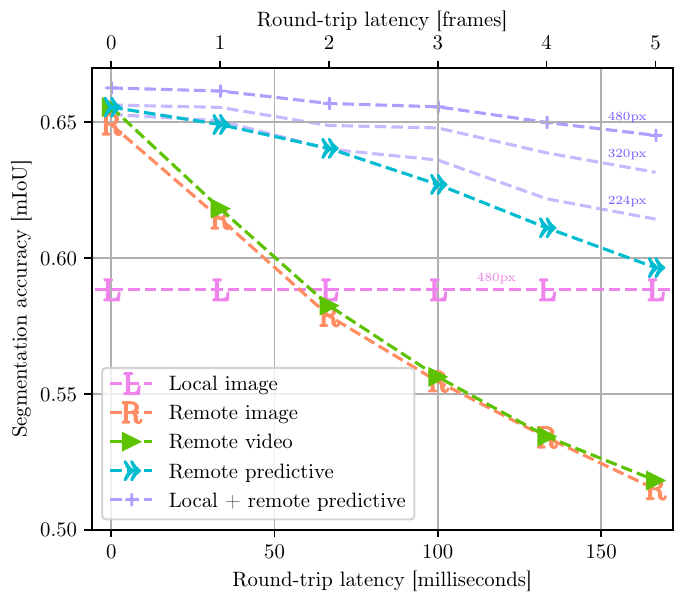}
    \end{center}
    \caption{
        Segmentation accuracy (mIoU) versus round-trip latency (milliseconds or frames).
        Further finetuned and evaluated on various local input resolutions.
    }
    \label{fig:miou_latency_localres}
\end{figure}

We evaluate performance across local input resolutions by finetuning the fused model (trained at 480 px) for 10 additional epochs at 224, 320, and 480 px.
The results are shown in~\cref{fig:miou_latency_localres}.
Our remote-assisted local model operates at far lower resolutions (e.g., 224 px) while achieving better accuracy under round-trip latency than other solutions.

\FloatBarrier

\subsection{Delay-input ablation}

We ablate the delay input by removing the delay conditioning from the remote model and finetuning for the same number of epochs.
This causes an average drop of 0.015 mIoU for local 224~px, showing that explicit delay inputs are beneficial.

\begin{table}[ht]
    \centering
    \caption{Delay-input ablation.}
    \scriptsize
    \setlength{\tabcolsep}{2.5pt}
    \begin{tabularx}{\linewidth}{@{}l | c c c c c c | c@{}}
        \toprule
        Experiment & \multicolumn{6}{c|}{Remote delay (frames)} & \\
        & 0 & 1 & 2 & 3 & 4 & 5 & Local \\
        \midrule[\heavyrulewidth]
        BDD 224px & 0.652 & 0.650 & 0.640 & 0.635 & 0.621 & 0.613 & --- \\
        BDD 224px (\textminus \ delay) & 0.622 & 0.628 & 0.627 & 0.625 & 0.615 & 0.603 & --- \\
        \bottomrule
    \end{tabularx}

    \label{tbl:delay_input_ablation}
\end{table}

\FloatBarrier

\subsection{Robustness to uplink compression}

We evaluate performance under x264/x265 compression (veryfast, zerolatency, bframes=0) to assess robustness to realistic streaming configurations.

\begin{table}[ht]
    \centering
    \caption{Robustness to uplink compression.}
    \scriptsize
    \setlength{\tabcolsep}{2.5pt}
    \begin{tabularx}{\linewidth}{@{}l | c c c c c c | c@{}}
        \toprule
        Experiment & \multicolumn{6}{c|}{Remote delay (frames)} & \\
        & 0 & 1 & 2 & 3 & 4 & 5 & Local \\
        \midrule[\heavyrulewidth]
        %
        %
        %
        BDD H.265 5 Mbps & 0.658 & 0.657 & 0.652 & 0.648 & 0.642 & 0.635 & --- \\
        BDD H.264 5 Mbps & 0.657 & 0.656 & 0.651 & 0.647 & 0.640 & 0.634 & --- \\
        BDD H.265 3 Mbps & 0.655 & 0.654 & 0.649 & 0.644 & 0.640 & 0.635 & --- \\
        BDD H.264 3 Mbps & 0.652 & 0.652 & 0.647 & 0.644 & 0.637 & 0.632 & --- \\
        BDD H.265 1 Mbps & 0.631 & 0.630 & 0.627 & 0.623 & 0.619 & 0.615 & --- \\
        BDD H.264 1 Mbps & 0.612 & 0.610 & 0.608 & 0.605 & 0.603 & 0.599 & --- \\
        \bottomrule
    \end{tabularx}

    \label{tbl:uplink_compression_ablation}
\end{table}

\FloatBarrier

\subsection{Compute and runtime breakdown}

\begin{table}[ht]
    \centering
    \caption{Compute and runtime breakdown by component.}
    \scriptsize
    \begin{tabularx}{\linewidth}{@{}l r r r r@{}}
        \toprule
        Component         & Params  & MACs      & GPU (ms) &  CPU (ms) \\
        \midrule[\heavyrulewidth]
        Remote ViT2D      & 42.206M & 199.683G  &   2.01   &   704.57  \\
        Delay embedding   &  0.264M &   0.0005G &   0.04   &     1.26  \\
        Remote ViT3D+head & 11.964M & 551.665G  &  24.38   & 34475.63  \\
        \midrule
        Local ViT2D+head  & 8.101M  &   1.455G  &   0.67   &     9.34  \\
        Additive fusion   & 0       &   0       &   0.04   &     0.01  \\
        \bottomrule
    \end{tabularx}

    \vspace{2pt}
    \tiny

    GPU: NVIDIA H100 fp16 PyTorch compiled \enspace | \enspace %
    CPU: Apple M3 Pro fp32 PyTorch 6 cores \enspace | \enspace %
    Local: 224px.

    \label{tbl:component_compute_runtime}
\end{table}

\FloatBarrier

\subsection{Additional dataset evaluation}

We train DeDelayed models on 35k clips from the first 229 sequences in the  Nymeria dataset~\cite{dataset_nymeria} and evaluate on 2k clips from 25 disjoint sequences with unseen environments and actions, using the same semantic segmentation task and ADE20K taxonomy.

\begin{table}[ht]
    \centering
    \caption{Generalizability to new environments and actions.}
    \scriptsize
    \setlength{\tabcolsep}{2.5pt}
    \begin{tabularx}{\linewidth}{@{}l | c c c c c c | c@{}}
        \toprule
        Experiment & \multicolumn{6}{c|}{Remote delay (frames)} & \\
        & 0 & 1 & 2 & 3 & 4 & 5 & Local \\
        \midrule[\heavyrulewidth]
        Nymeria~\cite{dataset_nymeria} & 0.2450 & 0.2451 & 0.2448 & 0.2446 & 0.2445 & 0.2443 & 0.2034 \\
        \bottomrule
    \end{tabularx}

    \label{tbl:nymeria_results}
\end{table}

\FloatBarrier

\subsection{Occlusion}

\begin{figure}[htbp]
    \centering
    \includegraphics[width=1.00\linewidth]{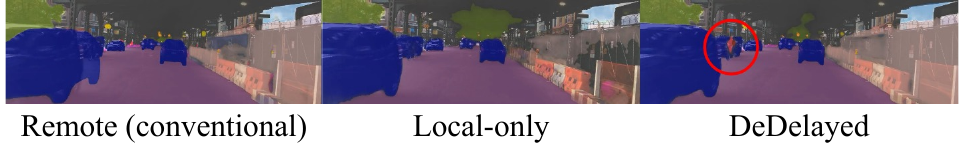}%
    \caption{
        Due to 167 ms RTT, the remote model has not yet seen any frames containing the recently unoccluded \textcolor{red}{cyclist}, so methods whose final inference is not local surely fail to detect it.
    }
    \label{fig:occlusion}
\end{figure}

\cref{fig:occlusion} shows a qualitative example of a newly unoccluded object missed by all methods except DeDelayed.

\FloatBarrier

\subsection{Sample validation pseudolabels}

\cref{fig:val_pseudolabels} visualizes 20 randomly selected samples from the validation set, with labels colored according to the conventional Cityscapes palette in \cref{fig:cityscapes_class_legend}.
The ego vehicle's hood is unlabeled, which is appropriate.
Core classes are generally accurate and present with low error: road, sky, vegetation, building, pole, person, car, truck, bus.
Most distant traffic lights and traffic signs are correctly identified.
In (row~2, column~1), the area beneath the fence is labeled sidewalk, which is debatable; a nearby non-traffic sign is left unlabeled, which is reasonable.
Some uncertain areas show speckles and jagged boundaries, often near partial occlusions.
Overall, the pseudolabels are close to human-annotated quality, with few exceptions.


\clearpage
\onecolumn
\FloatBarrier

\newcommand{\SubfigW}{0.25\linewidth}
\newcommand{\ImgW}{0.99\linewidth}
\newcommand{\RowSep}{-0.0ex}

\newcommand{\TilePrefix}{figures/dj_bdd_val_500/}
\newcommand{\TileSuffix}{_overlay_14.png}

\newcounter{tilecol}
\newcommand{\tile}[1]{%
  \begin{subfigure}{\SubfigW}
    \centering
    \includegraphics[width=\ImgW]{\TilePrefix#1\TileSuffix}
  \end{subfigure}%
  \stepcounter{tilecol}%
  \ifnum\value{tilecol}=4 \\[\RowSep]\setcounter{tilecol}{0}\fi
}

\begin{figure*}[htbp]
  \centering
  \setcounter{tilecol}{0}

  \tile{017}\tile{024}\tile{026}\tile{056}
  \tile{108}\tile{118}\tile{130}\tile{153}
  \tile{201}\tile{231}\tile{311}\tile{326}
  \tile{333}\tile{366}\tile{367}\tile{370}
  \tile{419}\tile{431}\tile{459}\tile{493}

  \caption{Randomly sampled validation pseudolabels.}
  \label{fig:val_pseudolabels}
\end{figure*}

\newcommand{\SwatchW}{2.0cm}
\newcommand{\SwatchH}{0.6cm}

\definecolor{unlabeled}{RGB}{0,0,0}
\definecolor{road}{RGB}{128,64,128}
\definecolor{sidewalk}{RGB}{244,35,232}
\definecolor{building}{RGB}{70,70,70}
\definecolor{wall}{RGB}{102,102,156}
\definecolor{fence}{RGB}{190,153,153}
\definecolor{pole}{RGB}{153,153,153}
\definecolor{trafficlight}{RGB}{250,170,30}
\definecolor{trafficsign}{RGB}{220,220,0}
\definecolor{vegetation}{RGB}{107,142,35}
\definecolor{terrain}{RGB}{152,251,152}
\definecolor{sky}{RGB}{70,130,180}
\definecolor{person}{RGB}{220,20,60}
\definecolor{rider}{RGB}{255,0,0}
\definecolor{car}{RGB}{0,0,142}
\definecolor{truck}{RGB}{0,0,70}
\definecolor{bus}{RGB}{0,60,100}
\definecolor{train}{RGB}{0,80,100}
\definecolor{motorcycle}{RGB}{0,0,230}
\definecolor{bicycle}{RGB}{119,11,32}

\newcommand{\sw}[2]{%
  \begingroup\setlength{\fboxsep}{0pt}%
  \colorbox{#2}{\parbox[c][\SwatchH][c]{\SwatchW}{\centering\scriptsize\textcolor{white}{#1}}}%
  \endgroup
}

\begin{figure*}[htbp]
    \centering
    \setlength{\tabcolsep}{0pt}
    \renewcommand{\arraystretch}{0.0}
    \begin{tabular}{ccccc}
        \sw{unlabeled}{unlabeled} &
        \sw{wall}{wall} &
        \sw{traffic sign}{trafficsign} &
        \sw{person}{person} &
        \sw{bus}{bus} \\
        \sw{road}{road} &
        \sw{fence}{fence} &
        \sw{vegetation}{vegetation} &
        \sw{rider}{rider} &
        \sw{train}{train} \\
        \sw{sidewalk}{sidewalk} &
        \sw{pole}{pole} &
        \sw{terrain}{terrain} &
        \sw{car}{car} &
        \sw{motorcycle}{motorcycle} \\
        \sw{building}{building} &
        \sw{traffic light}{trafficlight} &
        \sw{sky}{sky} &
        \sw{truck}{truck} &
        \sw{bicycle}{bicycle} \\
    \end{tabular}
    \caption{Cityscapes class palette.}
    \label{fig:cityscapes_class_legend}
\end{figure*}

\twocolumn

\end{document}